\documentclass[aps,pra,twocolumn,showpacs,nofootinbib,longbibliography,notitlepage]{revtex4-2}
\usepackage{etex}
\usepackage{amsmath,amssymb,amsthm}
\usepackage[colorlinks=true,citecolor=blue,urlcolor=blue]{hyperref}
\usepackage[pdftex]{graphicx}
\usepackage{times,txfonts}
\usepackage{braket}
\usepackage{color}
\usepackage{natbib}
\usepackage{amsmath,blkarray}
\usepackage{mathtools}
\usepackage{latexsym}
\usepackage{tabularx, booktabs}
\usepackage{graphics,epstopdf}
\usepackage{graphicx}
\usepackage{float}
\usepackage{graphicx}
\usepackage{amsfonts}
\usepackage{subcaption}
\usepackage{color,soul}
\newtheorem{theorem}{Theorem}

\newcommand{\be}{\begin{equation}}
\newcommand{\ee}{\end{equation}}
\newcommand{\ba}{\begin{eqnarray}}
\newcommand{\na}{\nonumber}
\newcommand{\ea}{\end{eqnarray}}

\newcommand{\tcr}{\textcolor{red}}

\begin{document}
	\title{ Device-independent full network nonlocality for  arbitrary-party and unbounded-input scenario }
		\author{Sneha Munshi}
	\author{ A. K. Pan }
	\email{akp@phy.iith.ac.in}
	\affiliation{Indian Institute of Technology Hyderabad, Kandi, Sangareddy, Telangana 502285, India}
	
	\begin{abstract}
 
 The nonlocality arising in a multi-party network involving multiple independent sources radically differs from the standard multipartite Bell nonlocality involving a single source. The notion of the full network nonlocality (FNN)  [\href{https://journals.aps.org/prl/pdf/10.1103/PhysRevLett.128.010403}{Phys. Rev. Lett.\textbf{128}, 010403 (2022)}] characterizes the quantum correlations that cannot be reproduced by a local-nonlocal model featuring one local source and the rest of nonlocal no-signaling sources. However, the demonstration of FNN was limited to bilocal and trilocal star-shaped network scenarios involving three or two dichotomic measurements for edge parties.  In this paper, we first demonstrate that a large class of prevailing network inequalities does not exhibit FNN. We then introduce an elegant set of arbitrary-party and unbounded-input network inequalities in star-shaped and linear-chain networks whose optimal quantum violation exhibits FNN, certifying that the nonlocality is genuinely distributed to the entire network.  Contrasting to existing demonstrations of FNN that inevitably require fixed-input and four-output elegant joint measurements for the central party, our generalized inequalities are more experimentally friendly, requiring only two-output measurements. Moreover, our derivation of optimal quantum violation is fully analytic and devoid of assuming the dimension of the quantum system, thereby showcasing its potential for device-independent self-testing. 

	\end{abstract}
	\maketitle

\section{Introduction} \textcolor{blue}{}Bell's theorem \cite{bell}, the epoch-making discovery in modern physics,  demonstrates the incompatibility of quantum theory with a local realist description of nature. This fundamental feature, widely known as Bell nonlocality \cite{brunnerreview}, is commonly manifested through quantum violation of a suitable Bell inequality, which provides the accreditation of device-independent self-testing of states and measurements \cite{mckague,mckague12,supicrev}. Besides providing the most
radical departure of quantum theory from the notion of classicality, Bell nonlocality constitutes a powerful toolbox for a plethora of applications, such as quantum cryptography \cite{acin07,Farkas2024,QCrev,Gras2021}, communication complexity \cite{burnum,burnum1} and certified random number generation \cite{Piro2010,Acin2012,Pivo2015}. 

A standard bipartite Bell experiment involves two distant parties.  The multipartite Bell scenario \cite{svet,uffink,mermin}, although hugely complex, is a natural extension of the bipartite scenario and features only one source that distributes a physical system to each of the parties. In contrast, the multipartite Bell experiment in a network configuration features multiple independent sources, leading to a novel form of multipartite nonlocality \cite{Bran2010,Cyril2012,Armi2014,Frit2012,Tavareview,Ker2019}, conceptually different from the standard multipartite Bell nonlocality. Lately, the network nonlocality has empowered a number of elegant certification protocols such as the falsification of the real quantum theory \cite{real,real2,Yao2024},   self-testing of all entangled states \cite{Acin2023}, and a set of commuting observables \cite{Sneha2023}.

The existing network topologies can be broadly divided into two categories; the closed and the open network. Although an open network inevitably requires multiple inputs per party, the closed networks exhibit nonlocality for a fixed input \cite{renou2019,Boreiri2023,Boreiri2024,Mao2024,Sekatski2023,Supic2020}. However, most open networks can be considered as a suitable composition of star-shaped  \cite{Armi2014,Tava2017,Andr2017,Sneha2021} and linear-chain networks \cite{Kaus2020,rahuladp}.  In recent times, the network nonlocality has been well explored by proposing various strategies leading towards the network inequalities \cite{Gisin2017,Chaves2016,Rosse2016,Tavakoli2016,Luo2018,TavakoliEJM,cont,Gisin2021,Gisi2020,Pozas2023,Camillo2024,Silva,Nava2020,snehaadp}. The experimental verification of some of the  proposals has also been reported \cite{Polino2023,Suprano2022}. One of the potential applications of network nonlocality is the future development of secure quantum internet \cite{Kimb2008,Wehner2018,Wang2022}. This demands a comprehensive understanding of the nonlocality in a network featuring a large number of parties and measurements. 
 
  In the study of network nonlocality, a pertinent question is whether the quantum violation of a network inequality warrants the distribution of nonlocal correlation to each party involved in the network. More precisely, whether the optimal quantum violation of a network inequality can be simulated if one of the sources produces local correlation. This question was recently raised in \cite{Kers2021} by introducing the concept of FNN and their proposed inequalities have recently been experimentally tested \cite{Gu2023,Wang2024,Huang2022}. 

To put things in perspective, let us consider a network topology featuring an arbitrary $n$ number of independent sources $S_k$ with $k\in[n]$ and introduce the following definition of the local nonlocal model (LNL) to address the question of simulability of the optimal quantum violation of a network nonlocality.

\textbf{Definition 1}: (LNL model) A model corresponding to a multi-source network is referred to as the LNL model if \textit{only one} source produces local correlations and the rest of the sources produce nonlocal no-signaling correlations.

 Clearly, in a LNL model, one source (say $S_{1}$) produces local correlation by sharing a physical state $\lambda$ with probability distribution $\mu(\lambda)$, and all other sources $S_{k\neq 1}$ produce nonlocal no-signaling correlations. As pointed out in \cite{Kers2021},  for the first network inequality \cite{Cyril2012} involving two independent sources, the nonlocality is \tcr{}\textit{not} distributed to each party, i.e., it admits a LNL model.  This necessitates introducing a precise notion of the nonlocal correlations,  fully distributed throughout the network, which is termed as FNN. Along the same line \cite{Kers2021}, we propose the following definition of FNN. 

\textbf{Definition 2:} 
 The network nonlocality is said to be FNN if the optimal quantum violation of a network inequality does not admit any LNL model, i.e., the nonlocality is \tcr{}genuinely distributed to the entire network.
 

In this paper, we first demonstrate that a large class of arbitrary-party and unbounded-input network inequalities does not exhibit FNN according to Definition 2. We then propose a simple but elegant set of unbounded input inequalities in star-shaped and linear-chain networks featuring an arbitrary number of parties such that the optimal quantum violation cannot be simulated by any LNL model. This, therefore, warrants the nonlocality being distributed to the entire network, exhibiting FNN. We stress that our derivation of the optimal quantum violation of the proposed network inequalities is fully analytical and devoid of assuming the dimension of the system, thereby enabling device-independent self-testing of states and measurements in a network.  

In contrast to \cite{Kers2021}, our scheme is generalized to an arbitrary number of parties and an unbounded number of inputs per party, and, crucially, does not require four-output elegant joint basis \cite{gisin} or Bell basis measurements. Since the elegant joint basis is quite challenging to implement, our scheme is thus more experimentally friendly. \tcr{}We discuss the efficacy of our demonstration of FNN compared to \cite{Kers2021}.
\begin{figure}[h]
\includegraphics[scale=0.44]{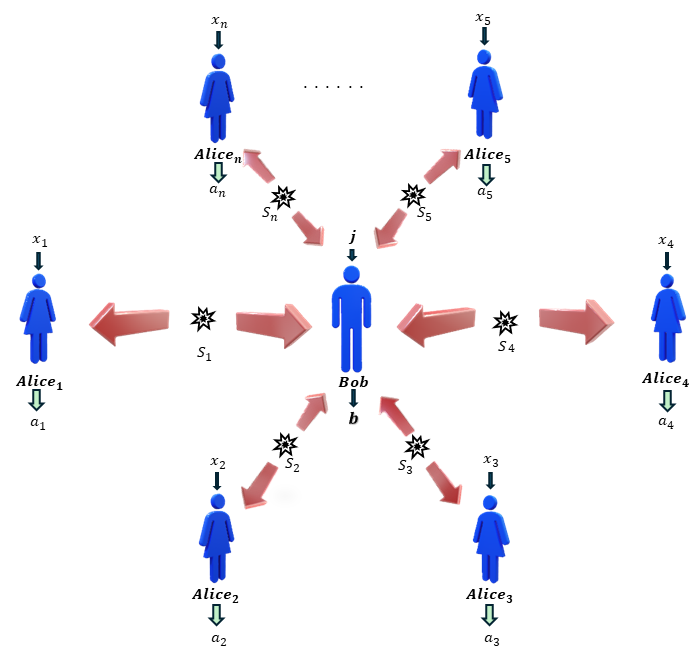}\caption{A star-shaped network featuring $n$ number of independent sources, each sharing a system with an edge party and the central party. }\label{FIG1}
				\end{figure}

   \section{Arbitrary source and unbounded input network scenario}\label{II} We first show that the optimal quantum violation of a large class of existing $n$-locality inequalities \cite{Cyril2012, Armi2014, Sneha2021} do not exhibit FNN. We start by considering the network inequalities in a star-shaped open network and then extend the argument to the linear-chain network in Appendix  \ref{A3}.   As depicted in Fig. \ref{FIG1}, the star-shaped network features an arbitrary $n$ number of independent sources $S_k$ and   $n$ number of edge parties (Alice$_{k}$) with $k\in[n]$. The central party (Bob) shares the physical system with each edge party  (Alice$_{k}$) emanating from the source  $S_{k}$. Further, Alice$_{k}$ (Bob) randomly performs an arbitrary $m$ ($2^{m-1}$) number of dichotomic measurements given by $A^k_{x_k}$ ($B_y$) respectively upon receiving the  inputs $x_{k}\in[m]$($y\in[2^{m-1}]$), producing the  output  $a_{k}(b)\in \{0,1\}$.

In a $n$-local model, the local hidden variables $\lambda_{k}$s, corresponding to the sources $S_{k}$s are assumed to be distributed according to the probability density function  $\mu_k(\lambda_{k})$ satisfying the relation $\int \mu_{k}{(\lambda_{k})} d\lambda_{k}=1$. Due to the source independence assumption, the joint distribution $\mu{(\lambda_{1},\lambda_{2},\cdots \lambda_{n} )}$ can be written in a factorized form 

	\ba
	\label{fac}
	\mu{(\lambda_{1},\lambda_{2},\cdots \lambda_{n})} =\prod\limits_{k=1}^n \mu_{k}{(\lambda_{k})}
	\ea 
	which is termed as the $n$-locality assumption \cite{Cyril2012}.  Using it, the  joint output probability  can be written as 
	\begin{eqnarray}\label{facn}
		&&P(a_{1}, a_{2}, \ldots , a_n, b,|x_{1},x_{2},\ldots x_n, y)\\
		\nonumber
	&=&\int\cdots \int \bigg(\prod\limits_{k=1}^n \mu_{k}{(\lambda_{k})}\hspace{3pt} d\lambda_{k}\hspace{3pt}  P(a_{k}|x_{k},\lambda_{k})\bigg) P(b|y,\lambda_{1},\lambda_{2}\ldots \lambda_{n}).
	\end{eqnarray}
	 Clearly,
	Alice$_{k}$'s outcome solely depends on the hidden variable $\lambda_{k}$, but Bob's outcome depends on all of the  $\lambda_{k}$s, where $k\in[n]$. In such a  scenario, a  generalized  $n$-locality inequality was proposed in \cite{Sneha2021} given by 
		\begin{equation}
		\label{deltapncnm}
		(\Delta^{n,m})_{n-l}=\sum\limits_{y=1}^{2^{m-1}} |{I}^{n,m}_{y}|^{\frac{1}{n}} \leq \alpha_{m}=\sum\limits_{q=0}^{\lfloor\frac{m}{2}\rfloor}\binom{m}{q}(m-2q)
		\end{equation} where $I_{y}^{n,m}$ is defined as 
		\begin{eqnarray}\label{Inmi}
I_{y}^{n,m}=\bigg\langle\prod\limits_{k=1}^{n}\sum\limits_{x_{k}=1}^{m}(-1)^{z^{y}_{x_k}} A_{x_{k}}^{k}B_{y}\bigg\rangle
		\end{eqnarray}

  Here $z_{x_k}^y$  takes a value of either $0$ or $1$, which is fixed by using the following encoding scheme adopted in \cite{Sneha2021,Ghorai2018}. Consider that   $z^{y}\in\{0,1\}^{m}$ is a  $m$-bit string with restriction of first bit to be $0$. Then, $z_{x_k}^y\in \{0,1\}$ is the $(x_{k})^{th}$ bit with $y\in \{1,2...2^{m-1}\}$.

Note that, the inequality in Eq. (\ref{deltapncnm}) reduces to the well-known bilocality inequality proposed by  Branciard \emph{et al.,} \cite{Cyril2012} for $m=n=2$, and the $n$-locality inequality proposed by Tavakoli \emph{et. al., }\cite{Armi2014} for $m=2$ and arbitrary $n$. 

 Using an efficient sum-of-squares (SOS) \cite{Sneha2021,pan21,snehachsh} approach and without using \tcr{}imposing any constraint on the dimension of the system, the optimal quantum violation of Eq. (\ref{deltapncnm}) is derived as $(\Delta^{n,m})_{Q}^{opt} =2^{m-1}\sqrt{m}$ \cite{Sneha2021}. Now, to address the question of simulability of  $(\Delta^{n,m})_{Q}^{opt}$ by a LNL model, we prove the following theorem.

\begin{theorem}
\label{th1}
The optimal quantum violation of  $n$-locality inequality in Eq. (\ref{deltapncnm}) admits a LNL model.
\end{theorem}
\emph{Proof:-} Following the  Definition 1, we consider that the source $S_1$ produces local correlation by sharing a physical state $\lambda$ between  Alice$_1$ and  Bob, whereas the other independent sources $S_{k\neq1},k\in[n]$ produce nonlocal no-signaling  correlations between Alice$_{k\neq 1}$ and Bob. Note that Bob holds  $n$ number of independent subsystems generated by the sources $S_k, k\in[n]$. Let us consider that  Bob performs the measurement corresponding to the input $y'=\{y_k\}\in\{1,2,\ldots,2^{m-1}\}^n$ where  $y_k\in[2^{m-1}]$ refers to the input corresponding to $k^{th}$ subsystem.  Further, by using  $(m-1)$ bit of local randomness, Bob fixes the order of implementation of $y_k$ such that $y_1=y_2=\cdots=y_n=y$ and hence Bob's input scenario  $y'=y\in[2^{m-1}]$. Now, each input $y_k$ produces the output $b^k\in\{0,1\}$ and hence Bob's output becomes    $b'=\{b^k\}\equiv \{0,1\}^n$.

Following the $n$-locality assumption,  the independence of the  sources $S_k,k\in[n]$ ensures the following factorized  joint  probability distribution 
\ba  &&\label{mnlpr}P(a_1,b',\textbf{a}|x_1,{y'},\textbf{x})=\\
\nonumber
&&\int d\lambda\mu(\lambda)P\left(a_1|x_1,\lambda\right)P\left(b^1|y_1,\lambda\right)\prod_{k=2}^n  P_{NS}\left(b^k,a_k|y_k,x_k\right)\ea
\textcolor{blue}{}where $\textbf{a}=a_2,a_3,\dots,a_n,$ and $\textbf{x}=x_2,x_3,\dots,x_n,$ and $\mu(\lambda)$ describes the probability density function  of the local variable $\lambda$. Here, the subscript $NS$ denotes the non-local no-signaling correlations. For each \textcolor{red}{}$k\in[n]\setminus\{1\}$, the source $S_k$  gives rise to the  nonlocal  no-signaling correlations, constrained by the following  conditions 
\ba \sum_{b^{k}}P_{NS}(b^k,a_k|y_k,x_k)=P_{NS}(a_k|x_k), 
\sum_{a_k}P_{NS}(b^k,a_k|y_k,x_k)=P(b^k|y_k)\hspace{0.75mm} \ea 
In the LNL model, the correlations can then  be written as
\ba \langle A^1_{x_1}\cdots A^n_{x_n}B^n_{{y'}}\rangle=\sum_{a_k, b^k } (-1)^{a_1+b^1+\sum_{k=2}^n{(a_k+b^k)}}P(a_1,b',\textbf{a}|x_1,{y'},\textbf{x})\ea
We suitably define  the probability $P_{NS}(b^k,a_k|y_k,x_k),$ for $ k\in[n]\setminus {1}$ for   our LNL model as follows
\textcolor{blue}{}\ba\label{nPRprev}P_{NS}(b^k,a_k|y_k,x_k)=\begin{cases}\frac{1}{2}, \quad z_{x_k}^{y_k}= b^k\oplus a_k,\\
0, \quad  \text{otherwise}
\end{cases} \ea

We consider that  Bob finally chooses his outcome as $b^1=b^2=\cdots =b^n=b$ such that $b'=b\in\{0,1\}, \forall k\in[n]$.  
Using Eq. (\ref{mnlpr}) and the correlations of Eq. (\ref{nPRprev}), we derive  the achievable maximum  value of $\Delta^{n,m}$ in the  LNL model   as   
 \ba
 (\Delta^{n,m})_{LNL}=(m2^{m-1})^{1-\frac{1}{n}}\left(\alpha_{m}\right)^{\frac{1}{n}}\ea
 
\begin{figure}[h]\begin{center}
\includegraphics[scale=0.52]{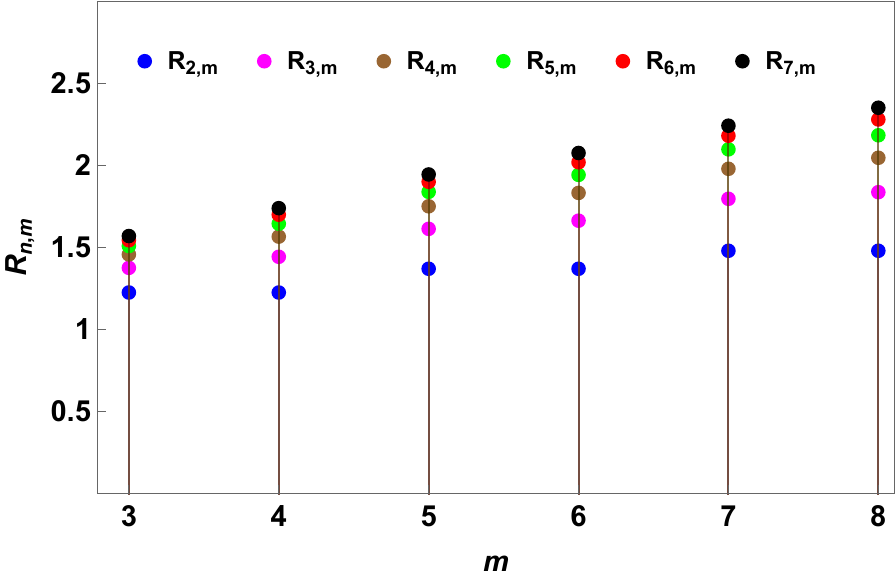}\caption{The figure illustrates the ratio $R_{n,m}$ for $m=3,4,\ldots 8$. For a fixed value of $m$, different color dots represent the value of the ratio $R_{n,m}$ for $n=2,3,\cdots, 7$. Note that for increasing values of $m$, the ratio $R_{n,m}$ and the difference between the points of the ratios  ($R_{n,m}$ for different $n$) increases. }\label{FIG2}
\end{center}
\end{figure}

To compare the LNL model and the optimal quantum value, we define the ratio $R_{n,m}=(\Delta^{n,m})_{LNL}/(\Delta^{n,m})_{Q}^{opt}$. For any arbitrary value of $n$, we find $R_{n,m}\geq 1$ irrespective of the value of $m$, as shown in Fig. \ref{FIG2}. For the asymptotic case of large $n$, we have $R_{n,m}\rightarrow \sqrt{m}$.  This concludes the proof. 

For $m= n=2$, we have $R_{n,m}= 1$ i.e.,  the optimal quantum violation of the bilocality inequality \cite{Cyril2012} does not warrant FNN,   as already pointed out in  \cite{Kers2021}. 

We note that the ratio $R_{n,m}$ increases monotonically with $n$ i.e., it is possible that for large $n$ cases, even if we increase the number of local sources in the LNL model,  still  $(\Delta^{n,m})_{Q}^{opt}$ remains simulable. To capture this fact, we define a more general local-nonlocal model, termed as $p$LNL model that features $p$ number of local sources, and $(n-p)$ number of nonlocal sources. Clearly, for  $p=1$, the model is equivalent to the LNL model by Definition 1. Using the correlations of  Eq. (\ref{nPRprev}), we derive the maximum value of $\Delta^{n,m}$ in a $p$LNL model to be 
\begin{equation}
 (\Delta^{n,m})_{pLNL }=(m2^{m-1})^{1-\frac{p}{n}}\left(\alpha_{m}\right)^{\frac{p}{n}} .  
\end{equation}
  It is straightforward to find the number of local sources $p$, for which  $(\Delta^{n,m})_{Q}^{opt}$ cannot be simulated by a $p$LNL model.
 For example, for $m=3$, $n=4$, we find $R_{3,4}^{p>2}<1$, i.e., the optimal quantum value $(\Delta^{4,3})_{Q}^{opt}$ cannot be simulated if more than two local sources are allowed.  

We note that in \cite{Sneha2023,snehaadp}, the network inequalities and their optimal quantum violation in the star-shaped network were derived by swapping the number of inputs so that Alice$_k$ and Bob perform $2^{m-1}$ and $m$ dichotomic measurements respectively. By following the above argument, we have checked that those set of inequalities are also incapable of exhibiting FNN.  Also, we consider another open network, the linear-chain one, and show that the optimal quantum violation of generalized network inequality \cite{rahuladp} is also inadequate to produce FNN. The detailed derivation is provided in the Appendix  \ref{A3}. This sets the motivation to formulate a new family of network inequalities for open networks that exhibit FNN.

\section{ Proposed $n$-locality inequalities exhibits FNN}  Consider the star-shaped network 
  in Fig.\ref{FIG1} where Alice$_{k} ( k\in[n]$)  performs the measurements of $m$ dichotomic observables denoted by $A^{k}_{x_{k}}$, according to the inputs $x_{k}\in[m]$  and obtains output  $a_{k}\in \{0,1\}$. However, the central party Bob measures the $m\times n$ number of dichotomic observables denoted by $B^{n,m}_{j,t}$ according to the inputs $(j,t)$ where $ j\in[m], t\in[n]$ and obtains the output $b\in \{0,1\}$. Taking the $n$-locality assumption in Eqs. (\ref{fac}) and (\ref{facn}), we propose the following $n$-locality inequality 
\begin{equation}
\label{Cnm}
(\mathcal{C}^{n,m})_{n-l}=\sum\limits_{t=1}^{n} \left(\sum\limits_{j=1}^{m}{I}^{n,m}_{j,t}\right)\leq 2mn-2n
\end{equation} where  $I_{j,t}^{n,m}$ is suitable linear combination of correlations defined  as
\ba I_{j,t}^{n,m}=\bigg\langle\prod\limits_{k\neq t}^{n}A^k_{x_k}(A^t_{x_t}+A^t_{x_t+1})B^{n,m}_{j,t}\bigg\rangle\ea 
 Here  $x_{k(t)}=j\in[m]$ and $A^{k}_{m+1}=-A^k_1$ and $t(\neq k)$ denotes that the observables corresponding to Alice$_t$.
 

 We derive the optimal quantum value as $\mathcal{C}^{n,m}$ is derived as \ba \label{mnq}(\mathcal{C}^{n,m})^{opt}_Q=2mn\cos\frac{\pi}{2m}\ea by using an efficient SOS approach and without assuming the dimension of the Hilbert space. The detailed derivation is deferred to the Appendix  \ref{A2}. Clearly, $(\mathcal{C}^{n,m})^{opt}_{Q}> (\mathcal{C}^{n,m})_{n-l}$ for any value of $m$ and $n$. Now, to address the question of FNN, that is, the simulability of $\mathcal{C}^{n,m})^{opt}_{Q}$ by any LNL model, we prove the following theorem.
\begin{theorem}\label{th2}
In a $n$-edge star-shaped network involving $m$-input per edge party, the simulation of $(\mathcal{C}^{n,m})^{opt}_Q$ in Eq. (\ref{mnq}) requires each edge party to produce a nonlocal correlation with the central party if the condition $n\leq   \left\lfloor\frac{1}{f(m)}\right\rfloor$ is satisfied where $ f(m)=m\left(1-\cos \frac{\pi}{2m}\right)$.
 \end{theorem}
 \emph{Proof:-} Consider a LNL model in a $n$-edge star-shaped network configuration featuring a local source $S_1$, as in Definition 1.  Since Bob has $n$ number of independent subsystems generated by the sources $S_k$s,  we assume that he performs the measurement corresponding to the input $(j',t)=\{j_k\}\times \{t\}\in\{1,2,\ldots,m\}^n\times\{1,2,\cdots n\}$ such that $(j_k,t)$ refers to the input corresponding to the $k^{th}$ subsystem, where $j_k\in[m]$ and a fixed value of $t\in[n]$.  Furthermore, using  $\log_2 mn$ bit of local randomness in Bob's subsystem,  he fixes the order of implementation of $(j_k,t)$ so that $j_1=j_2=\cdots=j_n=j$ where $j\in[m], t\in[n]$. Now for each input  $(j_k,t)$ corresponding to the $k^{th}$ subsystem, Bob produces the output $b^k\in\{0,1\}$ and hence his output becomes    $b'=\{b^k\}\equiv \{0,1\}^n$.
Note that the assumption of independence of the sources $S_k,k\in[n]$ leads to the factorization as in Eq. (\ref{mnlpr}).  For each $k\in[n]\setminus\{1\}$, Bob and Alice$_{k}$  give rise to the general no-signaling correlations, constrained by the following conditions
\begin{subequations} \begin{equation}
      \sum_{b^{k}}P_{NS}(b^k,a_k|(j_k,t),x_k)=P\big(a_k|x_k\big)
\end{equation}    
\begin{equation}\sum_{a_k}P_{NS}(b^k,a_k|(j_k,t),x_k)=P(b^k|(j_k,t))\end{equation} \end{subequations}

\textcolor{blue}{}which  holds for any fixed value of $t\in[n]$. The correlation terms can then be written as
\textcolor{blue}{}\ba\na \langle A^1_{x_1}\cdots A^n_{x_n}B^{n,m}_{j',t}\rangle=\sum_{a_k, b^k, k\in[n]} (-1)^{\sum_{k}{a_k+b^k}}P(a_1,b',\textbf{a}|x_1,(j',t),\textbf{x})\ea
where the probability $P_{NS}(b^k,a_k|(j_k,t),x_k)$ is  given by 
\ba\label{nPR}P_{NS}(b^k,a_k|(j_k,t),x_k\bigg)=\begin{cases}\frac{1}{2}, \quad j_k-x_k< (m-1), b^k\oplus a_k=0\\
\frac{1}{2}, \quad j_k-x_k=(m-1), b^k\oplus a_k=1\\
0, \quad  \text{otherwise}
\end{cases} \ea
Bob finally chooses his output as $b=b^k, \forall k\in[n]$  such that $b'=b\in\{0,1\}$.
Using the factorization in Eq. (\ref{mnlpr}) and the correlations in  Eq. (\ref{nPR}), we find that the   achievable maximum  value of  
 $(\mathcal{C}^{n,m})_{LNL}=(2mn-2)$. This  implies that the FNN is  exhibited when 
\ba\label{mn} n\leq  \left\lfloor\frac{1}{f(m)}\right\rfloor, \quad f(m)=m\left(1-\cos \frac{\pi}{2m}\right)\ea  
Hence for a given value of $m\geq 2$, there exists $n\in \mathbb{N}$ satisfying the relation in  Eq. (\ref{mn}) such that the optimal quantum violation of the $n$-locality inequality in Eq. (\ref{Cnm}) certifies that the nonlocality is distributed to the entire network. This completes the proof of FNN.

\emph{Example for $n=2$:-} For an arbitrary $m$ input  three-party $(n=2)$ \tcr{} network,  the $n$-local bound is $ (4m-4)$ and the optimal quantum value is $(\mathcal{C}^{2,m})_{Q}^{opt}=4m\cos{\frac{\pi}{2m}}$.
Using the above strategy, in the LNL model, the source $S_1$ ($S_2$) produces the local (nonlocal no-signaling) correlations, and thus we get $I_{j,1}^{2,m}=2, \forall j\in[m-1],I_{j=m,1}^{2,m}=0$ and $I_{j,2}^{2,m}=2, \forall j\in[m]$ which produce $(\mathcal{C}^{2,m})_{LNL}=(4m-2)$. Note that $(\mathcal{C}^{2,m})_{Q}^{opt}\geq (\mathcal{C}^{2,m})_{LNL}$ and \tcr{}the equality holds for $m=2$. Hence,  the optimal quantum value $(\mathcal{C}^{2,m})^{opt}_Q$ for $m>2$ requires that each edge party must generate nonlocal correlation with the central party, thus certifying the FNN.

\emph{Remark 1:-} It is evident from  Theorem \ref{th2} that for any arbitrary $n$-party star-shaped network, there exists   $m\in \mathbb{N}$ (number of inputs) such that  FNN can be exhibited. Alternatively, for any arbitrary inputs $m>2$, one always finds a suitable star-shaped network with $n$ edge parties to demonstrate FNN.

\emph{Remark 2:-} Motivated by the genuine nonlocality argument in the standard multipartite Bell scenario \cite{svet,uffink}, one may consider that Alice$_k$s $ (k\in[n]\setminus\{1\})$,  who share nonlocal no-signaling correlations with Bob, collaborate with each other. Such a scenario effectively reduces to the case of the tripartite network scenario, so that Alice$_1$ shares a local correlation, and the rest of the other parties share a nonlocal no-signaling correlation with Bob. It is straightforward to check that such a scenario also exhibits FNN for $m>2$.

\emph{Remark 3:-}We note here that due to the symmetric construction of the functional $\mathcal{C}^{n,m}$, the argument of FNN of our scheme remains invariant under the alteration of local and nonlocal resources. 

\emph{Remark 4:-} One may argue that the full power of a nonlocal no-signaling model is not being used to determine the FNN, and hence it is insufficient to simulate $(\mathcal{C}^{n,m})^{opt}_Q$. We argue that this is not the case here, as the nonlocal correlations we used provides the algebraic maximum value of the corresponding correlations. Therefore, we have exhausted the maximum nonlocal no-signaling correlations, and there cannot be any other LNL model that can produce a larger value than our derived value $(\mathcal{C}^{n,m})_{LNL}=2mn-2$. We provide explicit examples for  $n=2$ and $m=3$ in the Appendix  \ref{A1}.

  Note that in the Appendix  \ref{A4}, we have also proposed an unbounded-input network inequality in the linear-chain network that features arbitrary $n$ independent sources and $(n+1)$ parties and demonstrated that the optimal quantum violation of such network inequality exhibits FNN.  

 \section{Relation to an earlier work } Pozas-Kerstjens \emph{et al.,} \cite{Kers2021}  proposed two different forms of $n$-locality inequalities in star-shaped network. i) By considering two inputs per edge party, a $n$-locality inequality is derived that exhibits FNN \emph{only} for $n=3$. For any $n> 3$, the optimal quantum violation admits a LNL model.  ii) By using the inflation \cite{inflation} technique and computational method, they derived two tailored-made inequalities in the tripartite network scenario, exhibiting FNN. Note that for both forms of inequalities, the central party needs to perform the fixed-input measurements. For the quantum realization of the first case, a three-qubit entangled basis measurement is required, and for the second case, three-output elegant joint measurements or Bell state measurements are required. It is crucial to note that both three-qubit entangled basis measurements and two-qubit elegant joint measurements are extremely challenging to realize in the experiment. In \cite{Huang2022}, the elegant joint measurement is experimentally realized by using an additional ancilla degree of freedom and post-processing.    

 In contrast to \cite{Kers2021}, we formulated a simple,  elegant, and generalized set of $n$-locality inequalities in a star-shaped network featuring an arbitrary $n$ number of edge parties and an unbounded $m$ number of two-output measurements per edge party. Our work crucially differs from \cite{Kers2021} as instead of fixed-input and three-output measurements, here the central party performs $m\times n$ number of two-output measurements, therefore, making it more experimentally friendly. Further, the derivation of the optimal quantum violation of the $n$-locality inequalities is analytical and independent of any assumption on the dimension of the quantum system. We have also demonstrated the FNN for the arbitrary-party and unbounded-input linear-chain network.

\section{Summary and outlook} In sum, we demonstrated that a large class of prevailing  $n$-locality inequalities in star-shaped and liner-chain networks does not exhibit FNN, i.e., the optimal quantum violation can be simulated by a LNL model involving only one local source. We formulated a set of arbitrary-party and unbounded-input  $n$-locality inequalities that exhibit FNN. Our proposed inequalities are efficient and more experimentally friendly compared to the existing demonstration of FNN.  Also, the optimal quantum violation is derived without assuming the dimension of the quantum system by using an efficient SOS  technique. This feature enables device-independent self-testing of states and measurements in the network.

Our work paves the path for exploring the FNN for any arbitrary open network, as any such network can be considered as a composition of star-shaped and linear-chain networks. This eventually possesses potential applications for developing a secure quantum Internet. Our work thus opens up an exciting avenue for exploring FNN and noise-robust self-testing in complex networks.

\emph{Acknowledgment:-}SM acknowledges the support from the research grant  I-HUB/PDF/2022-23/06, Government of India. AKP acknowledges the support from the research grant SERB/MTR/2021/000908, Government of India.

\begin{widetext}
		\appendix

\section{Explicit demonstration of FNN for  three-party ($n=2$) star-shaped network for $m=3$}
\label{A1}
We first provide a detailed derivation of the optimal quantum violation of the network inequality for $m=3$ and $n=2$, by employing an efficient SOS approach. The same methods are used to derive the optimal quantum violation of the inequality Eq. (\ref{Cnm}). Further, we provide a detailed analysis of the FNN in this scenario.
\subsection{Derivation of optimal quantum value $(\mathcal{C}^{2,3})_{Q}^{opt}$}
By substituting $m=3$  and $n=2$ in Eq. (\ref{Cnm}), we get the following bilocality inequality
\begin{equation}
		\label{aC23}
		\mathcal{C}^{2,3}=\sum\limits_{t=1}^{2} \left(\sum\limits_{j=1}^{3}{I}^{2,3}_{j,t}\right)\leq 8 
		\end{equation} where  $I_{j,t}^{2,3}$ is a suitable linear combination of correlations defined  as follows.
  \ba \na I_{1,1}^{2,3}=\langle (A^1_{1}+A^1_{2})B^{2,3}_{1,1}A^2_1\rangle, \quad 
 I_{2,1}^{2,3}=\langle (A^1_{2}+A^1_{3})B^{2,3}_{2,1}A^2_2\rangle,  \quad 
 I_{3,1}^{2,3}=\langle (A^1_{3}-A^1_{1})B^{2,3}_{3,1}A^2_3\rangle,\\
 I_{1,2}^{2,3}=\langle A^1_1B^{2,3}_{1,2}(A^2_{1}+A^2_{2})\rangle,\quad  
 I_{2,2}^{2,3}=\langle A^1_{2}B^{2,3}_{2,2}(A^2_2+A^2_3)\rangle, \quad 
 I_{3,2}^{2,3}=\langle A^1_{3}B^{2,3}_{3,2}(A^2_3-A^2_1)\rangle.
 \ea

  To optimize $(\mathcal{C}^{2,3})_Q$, we consider  that $(\mathcal{C}^{2,3})_{Q}\leq \beta^{2,3}$  where $\beta^{2,3}$ is clearly the upper bound of $(\mathcal{C}^{2,3})_{Q}$.   This is equivalent to showing that there is a positive semidefinite operator $\langle \Gamma^{2,3}\rangle_{Q}\geq 0$ which can be expressed as $\langle \Gamma^{2,3}\rangle_Q=-(\mathcal{C}^{2,3})_{Q}+\beta^{2,3}$.	To prove this, we consider a set of suitable positive operators  $L^{2,3}_{j,t}$,   which are polynomial functions of   $A^1_{x_1}$, $A^2_{x_2}$, and  $B^{2,3}_{j,t}$ so that,
	\begin{align}
	\label{agamma1}
\langle\Gamma^{2,3}\rangle_{Q}=\sum\limits_{j=1}^3\sum\limits_{t=1}^2\frac{(\nu^{2,3}_{j})_{A_t}}{2}\langle\psi|(L^{2,3}_{j,t})^{\dagger}L^{2,3}_{j,t}|\psi\rangle
	\end{align} where $(\nu^{2,3}_{j})_{A_t}$s  are suitable positive numbers that will be specified soon. 
	We choose  
	\ba
	\label{anABC}\na
L^{2,3}_{1,1}|\psi\rangle_{A_1BA_2}&=&\left(\frac{{A}^1_{1}+{A}^1_{2}}{(\nu^{2,3}_{1})_{A_1}}\otimes A^2_1- B^{2,3}_{1,1}\right)|\psi\rangle_{A_1BA_2},\quad L^{2,3}_{2,1}|\psi\rangle_{A_1BA_2}=\left(\frac{{A}^1_{2}+{A}^1_{3}}{(\nu^{2,3}_{2})_{A_1}}\otimes A^2_2- B^{2,3}_{2,1}\right)|\psi\rangle_{A_1BA_2},\\
L^{2,3}_{3,1}|\psi\rangle_{A_1BA_2}&=&\left(\frac{{A}^1_{3}-{A}^1_{1}}{(\nu^{2,3}_{3})_{A_1}}\otimes A^2_2- B^{2,3}_{3,1}\right)|\psi\rangle_{A_1BA_2}, \quad
L^{2,3}_{1,2}|\psi\rangle_{A_1BA_2}=\left({A}^1_1\otimes\frac{{A}^2_{1}+{A}^2_{2}}{(\nu^{2,3}_{1})_{A_2}}-B^{2,3}_{1,2}\right)|\psi\rangle_{A_1BA_2},\\\na
L^{2,3}_{2,2}|\psi\rangle_{A_1BA_2}&=&\left({A}^1_2
		\otimes\frac{{A}^2_{2}+{A}^2_{3}}{(\nu^{2,3}_{2})_{A_2}}-B^{2,3}_{2,2}\right)|\psi\rangle_{A_1BA_2}, \quad L^{2,3}_{3,2}|\psi\rangle_{A_1BA_2}=\left({A}^1_3\otimes\frac{{A}^2_{3}-{A}^2_{1}}{(\nu^{2,3}_{3})_{A_2}}-B^{2,3}_{3,2}\right)|\psi\rangle_{A_1BA_2},
\ea
	with $(\nu^{2,3}_{1})_{A_t}=||({A}^t_{1}+{A}^t_{2})|\psi\rangle_{A_tB}||_{2}=\sqrt{2+\langle\{{A}^1_{1},{A}^1_{2}\}\rangle}$. Similarly $(\nu^{2,3}_{2})_{A_t}=\sqrt{2+\langle\{{A}^t_{2},{A}^t_{3}\}\rangle} $ and $(\nu^{2,3}_{3})_{A_t}=\sqrt{2-\langle\{{A}^t_{3},{A}^t_{1}\}\rangle}$, $\forall t\in[2]$.
	Again, in $\nu^{2,3}_1$, the superscript $2,3$ implies $n=2,m=3$,  i.e., the bilocal scenario ($n=2$) when  each edge party performs three $(m=3)$ measurements.  Substituting Eq. (\ref{anABC}) in   Eq. (\ref{agamma1}), we get 	\ba
 \label{agamma11}\langle\Gamma^{2,3}\rangle_{Q}=-(\mathcal{C}^{2,3})_{Q}         +\left(\sum\limits_{j\in[3],t\in[2]}(\nu^{2,3}_{j})_{A_t}\right)\ea 	
 
 It is evident from Eq. (\ref{agamma11}) that the optimal value of $(\mathcal{C}^{2,3})_{Q}$ is achieved if  $\langle \Gamma^{2,3}\rangle_{Q}=0$. This,  in turn,  provides,
	\ba
	\label{aSQopt}
	\nonumber(\mathcal{C}^{2,3})_{Q}^{opt}
	&=&\max\limits_{A^t_1,A^t_2,t\in[2]}\bigg[\sqrt{2+\langle\{{A}^1_{1},{A}^1_{2}\}\rangle}+\sqrt{2+\langle\{{A}^1_{2},{A}^1_{3}\}\rangle}+\sqrt{2-\langle\{{A}^1_{3},{A}^1_{1}\}\rangle}+\sqrt{2+\langle\{A^2_{1},A^2_{2}\rangle}+\sqrt{2+\langle\{A^2_{2},A^2_{3}\rangle}+\sqrt{2-\langle\{A^2_{3},A^2_{1}\rangle}\bigg]\ea
 We derive the optimal quantum value by using the following steps.
 \ba\na
(\mathcal{C}^{2,3})_{Q}^{opt}&\leq &\max\limits_{A^t_1,A^t_2,\in[2]}\bigg[\sqrt{3\left(6+\langle\{{A}^1_{1},{A}^1_{2}\}\rangle+\langle\{{A}^1_{2},{A}^1_{3}\}\rangle-\langle\{{A}^1_{3},{A}^1_{1}\}\rangle\right)}+\sqrt{3\left(6+\langle\{{A}^2_{1},{A}^2_{2}\}\rangle+\langle\{{A}^2_{2},{A}^2_{3}\}\rangle-\langle\{{A}^2_{3},{A}^2_{1}\}\rangle\right)}\bigg]\\
&=&\max\limits_{A^t_1,A^t_2,\in[2]}\bigg[\sqrt{3\left(6+\langle\{({A}^1_{1}+{A}^1_{3}),{A}^1_{2}\}\rangle-\langle\{{A}^1_{3},{A}^1_{1}\}\rangle\right)}+\sqrt{3\left(6+\langle\{({A}^2_{1}+{A}^2_{3}),{A}^2_{2}\}\rangle-\langle\{{A}^2_{3},{A}^2_{1}\}\rangle\right)}\bigg]\ea

For maximization, we need to  assume that $A^t_2=\frac{{A}^t_{3}+{A}^t_{1}}{(\nu_3)_{A_t}}$ where $(\nu_3)_{A_t}=\sqrt{2+\langle\{{A}^t_{1},{A}^t_{3}\}\rangle}$. This gives 
\ba\na
(\mathcal{C}^{2,3})_{Q}^{opt}&\leq &\max\limits_{A^t_1,A^t_2,\in[2]}\bigg[\sqrt{3\left(6+2(\nu_3)_{A_1}-\langle\{{A}^1_{3},{A}^1_{1}\}\rangle\right)}+\sqrt{3\left(6+2(\nu_3)_{A_2}-\langle\{{A}^2_{3},{A}^2_{1}\}\rangle\right)}\bigg]\\&=&\max\limits_{A^t_1,A^t_2,\in[2]}\bigg[\sqrt{3\left(6+2\sqrt{2+\langle\{{A}^1_{3},{A}^1_{1}\}\rangle}-\langle\{{A}^1_{3},{A}^1_{1}\}\rangle\right)}+\sqrt{3\left(6+2\sqrt{2+\langle\{{A}^2_{3},{A}^2_{1}\}\rangle}-\langle\{{A}^2_{3},{A}^2_{1}\}\rangle\right)}\bigg]\ea

Clearly, $(\mathcal{C}^{2,3})_{Q}^{opt}=6\sqrt{3}$ when $\langle\{{A}^t_{1},{A}^t_{3}\}\rangle=-1, \forall t\in[2]$. This gives that $(\nu_3)_{A_t}=1$ i.e., ${A}^t_{1}-{A}^t_{2}+{A}^t_{3}=0$. Using this relation, we get that   $\langle\{{A}^t_{1},{A}^t_{2}\}\rangle=\langle\{{A}^t_{2},{A}^t_{3}\}\rangle=1, \forall t\in[2]$. Also, 
 the condition $\langle \Gamma^{2,3}\rangle_{Q}=0$ yields 
$L_{j,t}^{2,3}|\psi\rangle_{A_1BA_2}=0;  \forall j\in[3],t\in[2]$ which provides the required measurement settings for  Bob.

It is important to note that the above  derivation is devoid of assuming the dimension of the system. However, the choices of observables required for optimal value can also be available for the local qubit system for each  edge party.  
\ba A^t_1=\sigma_z,\quad  A^t_2=\frac{\sqrt{3}\sigma_x+\sigma_z}{2}, \quad  A^t_3=\frac{\sqrt{3}\sigma_x-\sigma_z}{2}, \quad \forall t\in[2]\ea
Bob's two-qubit observables can be written as \ba \nonumber B^{2,3}_{1,1}=\frac{\sqrt{3}\sigma_x+\sigma_z}{2}\otimes \sigma_z ,\quad  B^{2,3}_{2,1}=\sigma_x\otimes \frac{\sqrt{3}\sigma_x+\sigma_z}{2}, \quad B^{2,3}_{3,1}=\frac{\sigma_x-\sqrt{3}\sigma_z}{2}\otimes \frac{\sqrt{3}\sigma_x-\sigma_z}{2}\\
 B^{2,3}_{1,2}=\sigma_z\otimes\frac{\sqrt{3}\sigma_x+\sigma_z}{2}  \quad B^{2,3}_{2,2}=\frac{\sqrt{3}\sigma_x+\sigma_z}{2}\otimes \sigma_x, \quad B^{2,3}_{3,2}=\frac{\sqrt{3}\sigma_x-\sigma_z}{2}\otimes \frac{\sigma_x-\sqrt{3}\sigma_z}{2} \ea
 Let us consider that the source $S_1(S_2)$ shares one copy of the two-qubit maximally entangled state $|\phi^+\rangle_{A_1B(BA_2)}=\frac{1}{\sqrt{2}}\big(|00\rangle+|11\rangle\big)$ with Alice$_1$ (Alice$_2$) and Bob which in turn provides that 
the joint state is given by 
\ba|\psi\rangle_{A_1BA_2}=|\phi^+\rangle_{A_1B}\otimes |\phi^+\rangle_{BA_2}\ea 
The above set of measurement settings and the state provide the optimal quantum violation  $(\mathcal{C}^{2,3})_Q^{opt}=6\sqrt{3}$.

\subsection{Proof of FNN in this scenario}
Now, let us consider the LNL model where the source $S_1$ shares a physical state $(\lambda)$ with Bob. Since, each observable is dichotomic, we get that $I_{1,1}^{2,3}+I_{2,1}^{2,3}+I_{3,1}^{2,3}=4$. The source $S_2$ generates nonlocal correlations between Bob and Alice$_2$ such that  for each  $t\in[2]$,   the  probability $P_{NS}(b_2,a_2|(j,t),x_2)$, satisfying no-signaling conditions, is   given by 
\ba P_{NS}(b_2,a_2|(j,t), x_2, )=\begin{cases}\frac{1}{2}, \quad j-x_2< 2, b\oplus a_2=0\\
\frac{1}{2}, \quad j-x_2=2, b\oplus a_2=1\\
0, \quad  \text{otherwise}
\end{cases} \ea

This strategy provides that
$I_{1,2}^{2,3}+I_{2,2}^{2,3}+I_{3,2}^{2,3}=6$ which is the algebraic maximum value of the correlations corresponding to the no-signaling source $S_2$. Thus we obtain the optimal value of $(\mathcal{C}^{2,3})_{LNL}$ to be $(\mathcal{C}^{2,3})_{LNL}=10<(\mathcal{C}^{2,3})_{Q}^{opt}$ which implies that the optimal quantum violation of the bilocality inequality in Eq. (\ref{aC23})  exhibits FNN.

Note that the no-signaling correlation provides the maximum algebraic value of the corresponding correlations which enables us to claim that the maximum possible value of the no-signaling correlations has been exhausted and no other LNL model can provide a better value for $(\mathcal{C}^{2,3})_{LNL}$.

\section{Derivations of the $n$-locality inequality in Eq. (\ref{Cnm}) and $(\mathcal{C}^{n,m})^{opt}_{Q}$ in Eq. (\ref{mnq})}\label{A2}

In Eq. (\ref{Cnm}),  for an arbitrary number of parties ($n$) and an unbounded number of inputs ($m$), the generalized   $n$-locality functional is defined as $\mathcal{C}^{n,m}=\sum\limits_{t=1}^{n} \left(\sum\limits_{j=1}^{m}{I}^{n,m}_{j,t}\right)
		$
where  \ba I_{j,t}^{n,m}=\bigg\langle\prod\limits_{k\neq t}^{n}A^k_{x_k}(A^t_{x_t}+A^t_{x_t+1})B^{n,m}_{j,t}\bigg\rangle\ea 
 where  $x_{k(t)}=j\in[m]$ and $A^{t}_{m+1}=-A^{t}_1, \forall t\in[n]$.   Since each observable is dichotomic, 
it is straightforward to show that  $(\mathcal{C}^{n,m})_{n-l}\leq 2mn-2n$.

Using the SOS  approach for optimization, we derive the optimal quantum violation   $(\mathcal{C}^{n,m})^{opt}_{Q}$ without assuming the dimension of the quantum system. Let us consider $(\mathcal{C}^{n,m})
^{opt}_{Q}\leq \beta^{n,m}$, where $\beta^{n,m}$ is the upper bound of $(\mathcal{C}^{n,m})_{Q}$.  This is equivalent to showing that there is a positive semidefinite operator $\langle \Gamma^{n,m}\rangle_{Q}\geq 0$ which can be expressed as $\langle \Gamma^{n,m}\rangle_{Q}=-(\mathcal{C}^{n,m})_{Q}+\beta^{n,m}$.	By invoking a set of suitable vectors $L^{n,m}_{j,t}|\psi\rangle$ which are polynomial functions of   $A^{k(t)}_{x_{k(t)}}$, $B^{n,m}_{j,t}$, $(x_{k(t)},j\in[m];  t,k\in[n])$, we can write
	\begin{equation}
	\label{amum}
	\langle\Gamma^{n,m}\rangle=\sum\limits_{t=1}^{n}\sum\limits_{j=1}^{m}\dfrac{(\nu^{n,m}_{j})_{A_{t}}}{2}\langle\psi|(L^{n,m}_{j,t})^{\dagger}L^{n,m}_{j,t}|\psi\rangle\hspace{5mm}
	\end{equation} and $(\nu^{n,m}_{j})_{A_{t}}$s are suitable positive numbers. The optimal quantum value of $(\mathcal{C}^{n,m})_{Q}$ is obtained if $\langle \Gamma^{n,m}\rangle_{Q}=0$, implying that 
	\begin{align}
	\label{aLnmi}
	L^{n,m}_{j,t}|\psi\rangle=0,\quad  \forall j\in[m], t\in[n]
	\end{align}
We consider  a set of suitable vectors $L^{n,m}_{j,t}|\psi\rangle$ as
	\ba
	\label{aLm}
L^{n,m}_{j,t}|\psi\rangle&=&\left(\prod\limits_{k\neq t}^{n}A^k_{x_k}\frac{(A^t_{x_t}+A^t_{x_t+1})}{(\nu^{n,m}_{j})_{A_{t}}}
		 -B^{n,m}_{j,t}\right) |\psi\rangle
		\ea
	where $(\nu^{n,m}_{j})_{A_{t}}=||({A}^{t}_{x_t}+{A}^{t}_{x_t+1})|\psi\rangle_{A_tB}||_{2}=\sqrt{2+\langle\{A^{t}_{x_t},A^{t}_{x_t+1}\}\rangle}$, for each $t\in[n]$; $|\psi\rangle=\otimes_{t=1}^n|\psi\rangle_{A_tB}$ and $|\psi\rangle_{A_tB}$ is the state being shared by the source $S_t$.         
	Putting 	$L^{n,m}_{j,t}|\psi\rangle$ of Eq. (\ref{aLm}) in Eq. (\ref{amum}), we get 	\ba\label{amuC}\langle\Gamma^{n,m}\rangle_{Q}=-(\mathcal{C}^{n,m})         +\sum\limits_{t=1}^{n}\sum\limits_{j=1}^{m}(\nu^{n,m}_{j})_{A_t}\ea
 Since $\langle\Gamma^{n,m}\rangle_{Q}\geq 0$, we have 
	\begin{eqnarray}\label{amuC1}(\mathcal{C}^{n,m})_{Q}^{opt}&=& \max_{A_t }\bigg[\sum\limits_{t=1}^{n}\sum\limits_{j=1}^{m}(\nu^{n,m}_{j})_{A_t}\bigg]=2mn\cos\frac{\pi}{2m}.\ea

{\bf{Example: (Derivation for $m=5$):}}  The dimension-independent optimization of  $\sum\limits_{j=1}^{m}(\nu^{n,m}_{j})_{A_t}$ for  $m=2,3,4$ is  provided in \cite{snehachsh}. For larger values of $m$,  the calculation is cumbersome.  For better understanding, we provide the detailed dimension-independent optimization for $m=5$, i.e., we optimize $\left(\sum\limits_{j=1}^5(\nu^{n,5}_j)_{A_t}\right)$. Using the norm as defined earlier and  the inequality $\sum\limits_{j=1}^{m}f_j\leq \sqrt{m\sum\limits_{j=1}^{n}(f_j)^2}$, ($\forall f_j\geq 0 $), we can write 
\ba
\left(\sum\limits_{j=1}^5(\nu^{n,5}_j)_{A_t}\right)&\leq&\sqrt{5\sum\limits_{j=5}^5\left[(\nu^{n,5}_j)_{A_t}\right]^2}
 =\sqrt{5\bigg[10+ \tau_5 \bigg]}
 \ea
  where \ba\tau_5 =\langle\{A^t_2,(A^t_1+A^t_3)\}\rangle+\langle \{A^t_4,(A^t_3+A^t_5)\}\rangle-\langle\{A^t_1,A^t_5\}\rangle\ea
 For maximization, by considering $A^t_2=(A^t_1+A^t_3)/\nu_5,$and  $ A^t_4=(A^t_3+A^t_5)/\nu'_5 $, we get
 \ba\na\tau_5=2\sqrt{2+\langle\{A^t_1,A^t_3\}\rangle}+2\sqrt{2+\langle\{A^t_3,A^t_5\}\rangle}-\langle\{A^t_1,A^t_5\}\rangle
\leq2\sqrt{2\bigg(4+\langle\{(A^t_1+A^t_5),A^t_3\}\rangle\bigg)}-\langle\{A^t_1,A^t_5\}\rangle
 \ea
By considering   $A^t_3=(A^t_1+A^t_5)/\nu''_5$, we see \ba\na\tau_5&\leq & 2\sqrt{2\bigg(4+2\sqrt{2+\langle\{A^t_1,A^t_5\}\rangle}\bigg)}-\langle\{A^t_1,A^t_5\} \ea
 
 The maximization of $\tau_5$ provides the optimization  condition  $\{A^t_1,A^t_5\}=-(\sqrt{5}+1)/2$ which implies $\nu''_5=\sqrt{2+\{A^t_1,A^t_5\}}=(\sqrt{5}-1)/2$.  Also, we have found $\{A^t_1,A^t_3\}=\{A^t_3,A^t_5\}=\{A^t_2,A^t_4\}=-\{A^t_1,A^t_4\}=-\{A^t_2,A^t_5\}=(\sqrt{5}-1)/2, \quad \{A^t_1,A^t_2\}=\{A^t_2,A^t_3\}=\{A^t_3,A^t_4\}=\{A^t_4,A^t_5\}=(\sqrt{5}+1)/2,$ and $ \{A^t_1,A^t_4\}=\{A^t_2,A^t_5\}=-(\sqrt{5}-1)/2$.   Consequently, we get   $\nu_5=\nu'_5=(\sqrt{5}+1)/2$ , and $(\nu^{n,5}_j)_{A_t}=\sqrt{ (5 + \sqrt{5})/2}, \forall j\in[5],t\in[n]$. Putting all the above values, we finally get 
\ba\left(\sum\limits_{j=1}^5(\nu^{n,5}_j)_{A_t}\right)=10\cos\frac{\pi}{10}\ea

For an arbitrary $n$, we get $(\mathcal{C}^{n,5})_{Q}^{opt}= 10n\cos\frac{\pi}{10}$, which is in accordance with Eq. (\ref{amuC1}).


\section{linear-chain network: Non-FNN of existing  inequalities}\label{A3}

\begin{figure}[ht]
			{{\includegraphics[width=0.93\linewidth]{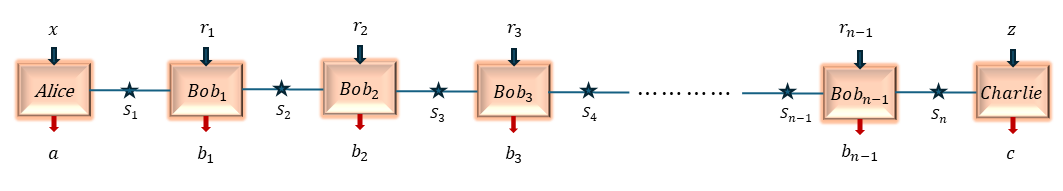}}}
			\caption{Linear-chain network featuring $(n+1)$ parties and $n$ independent sources }
			\label{achainn}
		\end{figure}

The  linear-chain network    features an $n$ number of sources $S_{\kappa\in[n]}$,  arbitrary  $(n+1)$ number of parties,  Alice, Bob$_{k\in[n-1]}$ and Charlie. The source $S_{\kappa\in[n]}$ shares the physical system with two adjacent parties. This implies that for $\kappa=1$ and $n$, the source shares a system with Alice-Bob$_1$ and Bob$_{n-1}$-Charlie, respectively. However, for the rest of the sources $S_{\kappa}, \kappa\in\{2,3,\cdots n-1\}$, a system is shared between two consecutive Bobs, i.e., Bob$_{k}$ and Bob$_{k+1}$ with $ \forall k\in[n-2]$.  Alice (Charlie)  performs the measurements of $m$ dichotomic observables denoted by $A_{x}(C_z)$, according to the inputs $x(z)\in[m]$  and obtains output  $a(c)\in \{0,1\}$. However, the middle  parties Bob$_{k\in[n-1]}$ perform the measurements of $2^{m-1}$ number of dichotomic observables denoted by $B^{k}_{r_k}$, according to the inputs  $r_k\in[2^{m-1}]$  and obtains output  $b_k\in \{0,1\}$. 
 The generalized unbounded-input  $n$-locality inequality in    linear-chain network  is given by \cite{rahuladp} 
		\begin{equation}
		\label{adeltapncnm}
		\mathcal{I}^{n,m} =\sum\limits_{r_k=1}^{2^{m-1}} \sqrt{|{J}^{n,m}_{r_k}|}\leq\sum\limits_{q=0}^{\lfloor\frac{m}{2}\rfloor}\binom{m}{q}(m-2q) \equiv \alpha_m
		\end{equation} where we define $J_{r_k}^{n,m}$ as suitable linear combination of correlations, given by 
		\begin{eqnarray}\label{aInmi}
J^{n,m}_{r_k}=\bigg\langle\sum_{x=1}^{m}(-1)^{\theta_{x}^{r_k}} A_{x}\otimes_{k=1}^{n}B^k_{r_k}\sum_{z=1}^{m}(-1)^{\theta_{z}^{r_k}} C_{z}\bigg\rangle
		\end{eqnarray}
		Here, the value of $\theta^{r_k}_x(\theta^{r_k}_z)$ is fixed by following the similar encoding scheme used for $z_{x_k}^{y}$ in Eq. (\ref{Inmi}) of Sec \ref{II}. Note that for $n=2$, it again reduces to the well-known tripartite bilocal network scenario \cite{Cyril2012}. The optimal quantum value can be derived using the SOS technique which gives  $(\mathcal{I}^{n,m})_{Q}^{opt}=2^{m-1}\sqrt{m}$.

In a LNL model, one assumes that the source $S_1$ is sharing a physical state $\lambda$ with Alice and Bob$_1$ where the other sources $S_{\kappa\neq 1}$ are producing nonlocal correlations, restricted by the no-signaling conditions. 
The assumption of   independence  of the sources $S_{\kappa},\kappa\in[n]$ ensures the factorization of the joint probability distribution $P(a,\textbf{b},c|x,\textbf{r},z)$ as follows:
  \textcolor{blue}{}\ba  \label{achpr}P(a,\textbf{b},c|x,\textbf{r},z)&=&\int d\lambda\hspace{1mm}\mu(\lambda)P\left(a|x,\lambda\right)P\left(b^{1}_1|r^{1}_{1},\lambda\right)\left(\prod_{k=1}^{n-2}  P_{NS}(b^{2}_k,b^{1}_{k+1}|r^{2}_{k},r^{1}_{k+1}) \right)P_{NS}\left(b^2_{n-1},c|r^2_{n-1},z\right)\ea
\textcolor{blue}{}where $\textbf{b}=\{b_k^1,b_k^2\},\textbf{r}=\{r_k^1,r_k^2\},  \forall k\in[n-1]$ and  the local hidden variable follows the probability density function $\mu(\lambda)$. Here, we consider that  Bob$_k$ posses $1$ bit of local randomness to fix the order of implementation of two inputs such that $r^{1}_k=r^{2}_k=r_k, \forall k\in[n-1]$,  and performs the measurement $B^k_{r_k}$ corresponding to the input $r_k\in[2^{m-1}]$. He finally chooses the output $b^1_k=b^2_k=b_k,\forall k\in[n-1]$. However, for each $k$, Bob$_k$ and Bob$_{k+1}$  give rise to general no-signaling correlations, constrained by the following conditions. \textcolor{blue}{}\ba\label{ansch} \sum_{b_{k}(b_{k+1})} P_{NS}(b_k,b_{k+1}|r_k,r_{k+1})&=&P\bigg(b_k(b_{k+1})\big|r_k(r_{k+1})\bigg) \ea
\textcolor{blue}{}
\textcolor{blue}{}where the probability  is given by
\textcolor{blue}{}\ba\label{anPRprev}P_{NS}(b_k,b_{k+1}|r_k,r_{k+1})=\begin{cases}\frac{1}{2}, \quad  b_k\oplus b_{k+1}=0,\\
0, \quad  \text{otherwise}
\end{cases} \ea

The source $S_n$ produces nonlocal no-signaling correlations following the Eq. (\ref{ansch}) such that the probability is given by 

\ba\label{anPRch}P_{NS}(b_{n-1},c|r_{n-1},z)=\begin{cases}\frac{1}{2}, \quad \theta^{r_{n-1}}_{z}= b_{n-1}\oplus c,\\
0, \quad  \text{otherwise}
\end{cases} \ea

Following the factorization in Eq. (\ref{achpr}) and the correlations in  Eqs. (\ref{anPRprev}-\ref{anPRch}), the maximum achievable value of  
 $(\mathcal{I}^{n,m})_{LNL}$ becomes    $\sqrt{m2^{m-1}\alpha_m}$. This then implies that for any value of $n$ and $m$, one can always simulate the optimal quantum value in an LNL model featuring a single local source. Therefore, the above nonlocal correlations in a linear chain network do not exhibit   FNN. This motivates us to construct a class of inequalities for an arbitrary-party and unbounded input linear chain network, that exhibits FNN.

 \section{ FNN in linear-chain network}
\label{A4}
 
 For the above linear-chain topology, let us consider that Alice and Charlie choose their measurement settings $A_x$ and $C_z$ upon receiving the inputs $x,z\in[m]$ and obtain the output $a,c\in\{0,1\}$. However, Bob chooses the measurements $B^{k}_{r_k,t}$ upon receiving the inputs $r_k\in[m],t\in[2]$, implying that each Bob$_k$ has a total choice of measurements $2m$, producing  the output $b_{k,t}\in\{0,1\}$.
  We propose the following $n$-locality inequality.

 \begin{equation}
		\label{adeltapncnm}
		\mathcal{T}_{n,m} =\sum\limits_{r_k=1}^{m} {l}^{n,m}_{r_k,1}+\sum\limits_{r_k=1}^{m} {l}^{n,m}_{r_k,2}\leq 4m-4
		\end{equation} where we define $l_{r_k,1}^{n,m}$ and $l_{r_k,2}^{n,m}$ as suitable linear combination of correlations, given by 
		\begin{eqnarray}\label{alnmi}\na
{l}^{n,m}_{r_k,1}=\bigg\langle\sum_{x=1}^{m}(A_x+A_{x+1})\otimes _{k=1}^nB^{k}_{r_k,1}C_z\bigg\rangle, \\ {l}^{n,m}_{r_k,2}=\bigg\langle\sum_{z=1}^{m}A_x\otimes _{k=1}^nB^{k}_{r_k,2}(C_z+C_{z+1})\bigg\rangle\end{eqnarray}
where $A_{m+1}(C_{z+1})=-A_1(-C_1)$.  Using the similar SOS approach as used before, we  derive the optimal quantum value $(\mathcal{T}^{n,m})^{opt}_Q=4m\cos\frac{\pi}{2m}$.\\

Now, let us consider a LNL model with a similar strategy (that exhausts all the possible LNL models by providing maximum possible algebraic value) as described above. Following the $n$-locality assumption,   the joint probability distribution can be written as 
  \textcolor{blue}{}\ba  \label{achprnew}P(a,\textbf{b}_t,c|x,\textbf{r}_t,z)&=&\int d\lambda\hspace{1mm}\mu(\lambda)P\left(a|x,\lambda\right)P\left(b_{1,t}^1|r^{1}_{1,t},\lambda\right)\left(\prod_{k=1}^{n-2}  P(b^{2}_{k,t},b^{1}_{k+1,t}|r^{2}_{k,t},r^{1}_{k+1,t}) \right)P\left(b^{2}_{n-1,t},c|r^{2}_{n-1,t},z\right)\ea
where $\textbf{b}_t=\{b_{k,t}\}$ and $\textbf{r}_t=\{r_{k,t}\},  \forall k\in[n-1]$ and a fixed $t\in[2]$.  Each Bob$_k$  posses $1$ bit of local randomness to fix the order of implementation of two inputs, such that, $r^{1}_{k,t}=r^{2}_{k,t}=r_{k,t}, \forall k\in[n-1],t\in[2]$,  and finally fixes the output as $b^{1}_{k,t}=b^{2}_{k,t}=b_{k,t}$.
The source $S_n$ produces nonlocal no-signaling correlations as in Eq. (\ref{ansch}) and   for any $t\in[2]$, the probability is given by 
\ba\label{anPRchg}P_{NS}(b_{n-1,t},c|r_{n-1,t},z)=\begin{cases}\frac{1}{2}, \quad r_{n-1,t}-z< (m-1),\quad  b_{n-1,t}\oplus c=0\\
\frac{1}{2}, \quad r_{n-1,t}-z=(m-1), \quad b_{n-1,t}\oplus c=1\\
0, \quad  \text{otherwise}
\end{cases} \ea
 
Using  Eq. (\ref{achprnew}) and the correlations in Eqs. (\ref{anPRprev}) and  (\ref{anPRchg}), we obtain the maximum achievable value $(\mathcal{T}^{n,m})_{LNL}=(4m-2)<(\mathcal{T}^{n,m})^{opt}_Q$, for any value of $m>2$. This then  implies that the network nonlocality is distributed over the whole network, exhibiting  FNN. \\

  \end{widetext}

\end{document}